\documentclass[pre,aps,showkeys,nofootinbib,preprint]{revtex4-1}
\usepackage{graphicx}
\usepackage{setspace}
\begin{document}
\newcommand{\be}{\begin{equation}}
\newcommand{\ee}{\end{equation}}
\newcommand{\bea}{\begin{eqnarray}}
\newcommand{\eea}{\end{eqnarray}}

\title{Microcanonical Origin of the Maximum Entropy Principle for Open Systems}
\author{Julian Lee\footnote{E-mail: jul@ssu.ac.kr.}}
\affiliation{
\vspace{0.7cm}
Department of Bioinformatics and Life
Science, Soongsil University, Seoul 156-743, Korea\\
}

\begin{abstract}
The canonical ensemble describes an open system in equilibrium with a heat bath of fixed temperature. The probability distribution of such a system, the Boltzmann distribution, is derived from the uniform probability distribution of the closed universe consisting of the open system and the heat bath, by taking the limit where the heat bath is much larger than the system of interest. Alternatively, the Boltzmann distribution can be derived from the Maximum Entropy Principle, where the Gibbs-Shannon entropy is maximized under the constraint that the mean energy of the open system is fixed. To make the connection between these two apparently distinct methods for deriving the Boltzmann distribution, it is first shown that the uniform distribution for a microcanonical distribution is obtained from the Maximum Entropy Principle applied to a closed system. Then I show that the target function in the Maximum Entropy Principle for the open system,  is obtained by partial maximization of Gibbs-Shannon entropy of the closed universe over the microstate probability distributions of  the heat bath. Thus, microcanonical origin of the Entropy Maximization procedure for an open system, is established in a rigorous manner, showing the equivalence between apparently two distinct approaches for deriving the Boltzmann distribution. By extending the mathematical formalism to dynamical paths, the result may also provide an alternative justification for the principle of path entropy maximization as well.
\end{abstract}
\maketitle
\section{Introduction}
Although canonical ensemble is widely used for computing various thermodynamic quantities, the most fundamental definitions and postulates in statistical mechanics are all formulated for a microncanonical ensemble, which describes a system in isolation. In contrast, the canonical ensemble describes a system in thermal contact with a heat bath with a fixed temperature. This ensemble can be considered as a special limit of the microcanonical ensemble, where the system consists of weakly coupled two subsystems and one of them is much larger than the other. Therefore, one might say that the microcanonical formulation of statistical physics is more fundamental, and the canonical formalism can be derived from the former. 

The basic postulate of the statistical mechanics for an isolated system, with fixed parameters specifying the macroscopic state, is that all the microscopic states consistent with these macroscopic parameters have equal probability of being occupied by the system\cite{PB,Rei,Path}. The Boltzmann distribution, the probability distribution of microstates of canonical ensemble,  is then obtained by dividing the system into the system of interest and the heat bath, the latter being much larger than the former, and then summing over all the microstates of the latter\cite{PB,Path,Chan}.

An alternative derivation of the Boltzmann distribution is from the so-called Maximum Entropy principle\cite{PB,Rei,jaynes57,jaynes03}. Here the Gibbs-Shannon entropy $H= -\sum_i p_i \ln p_i$ is maximized under the constraint that the expectation value of energy, $\sum_i p_i E_i$, is fixed to a certain value, where $p_i$ and $E_i$ are the occupation probability and the energy of the  microstate labeled as $i$. No transparent connection between the two approaches has been established, especially  why fixing the temperature of the heat bath is equivalent to fixing the mean energy of the open system.  In fact, the temperature of the heat bath determines the most probable value of open system energy, which is determined by the condition that  the derivative of the system entropy with respect to energy should be equal to the inverse of the heat bath temperature.  In the limit of infinite system size,   the relative fluctuation of energy from the expectation value becomes tiny, and the   expectation value of energy becomes  essentially the same as the most probable value of energy, but the equivalence breaks down once we consider an open system with a small size. In this era where small systems with nano-scale sizes can be probed with ease and mesoscopic systems have become subject of interest\cite{bus02}, it is important to formulate statistical physics without taking the limit of infinite system size. In this work, only the size of the heat bath is taken to infinity by definition, but the system size will be kept finite. 

Providing the microcanonical origin for the Maximum Entropy principle is important because the latter is applied also to dynamical systems\cite{FK,jaynes85,Stock,philken,Steve11,HG,CecileJSM}. In this case, the probability appearing in the Gibbs-Shannon entropy is the probability of each dynamical path, and the entropy is maximized under the constraint that the expectation value of quantities such as flux is held fixed, to obtain Boltzmann-like distribution for the path probability. Since the mathematics of the present work can be reinterpreted in the context of dynamical systems, an alternative justification for such a path probability distribution can be provided, which does not have as a firm foundation as the Boltzmann distribution in the case of the equilibrium system. 

In this work, I show that the Maximum Entropy principle for the open system simply follows from the Maximum Entropy principle applied to the closed universe consisting of the open system and the heat bath, by performing partial maximization of the Gibbs-Shannon entropy of the closed universe, over the probability distributions of microstates of the heat bath. 
The organization of the paper is as follows. I first explain the entropy maximization for the closed system, and how uniform distribution is obtained as the result.  The standard derivation of Boltzmann distribution from the uniform distribution by dividing the closed universe into the open system and the heat bath, is also briefly reviewed.
It is then shown that the target function to be used for entropy maximization procedure of an open system, the Gibbs-Shannon entropy plus the Lagrange term constraining the mean energy  of the open system,    is derived by performing a partial maximization of the Gibbs-Shannon entropy of the closed universe, over the probability distribution of microstates of the heat bath.  Therefore, the microcanonical origin of the entropy maximization procedure for an open system is established, showing that  equilibrating an open system with a heat bath of fixed temperature is equivalent to constraining its mean energy. Thus, two apparently distinct ways of deriving the Boltzmann distribution are shown to be the same thing.

\section{Maximum Entropy Principle for Close System and the Uniform Distribution}
The basic postulate of the statistical mechanics is that all the microstates, consistent with the given macroscopic parameters, have the same probability of being occupied by the system.  For concreteness,  let us suppose that the total energy $E$ of the system is given. Then the uniform distribution for the microcanonical ensemble describing this system is expressed as:
\be
p_i = \frac{\delta_{E, E_i}}{\Omega(E)} \label{uni}
\ee
where  $\Omega(E)$ is the total number of microstates with energy $E$. It is to be understood that  all the other parameters such as volume or particle number are also fixed, and  states with any other values of these parameters are excluded when counting the state. Only the constraint of fixed energy, which is of interest here, will be explicitly treated with the Kronecker delta  function\footnote{For notational simplicity, we assume discrete energy levels in this work.  When the energy level is continuous, it is more natural to count the number of states with energy values lying between $E$ and $E+dE$ with some small number $dE$, and $\delta_{E, E_i}$ in Eq.(\ref{uni}) should be replaced by $\theta(E_i-E) \theta(E+dE - E_i)$, with corresponding minor modifications of the mathematics in the following sections.}, and any other macroscopic parameters will be suppressed throughout the manuscript for notational simplicity.  

Rather than being treated as a starting assumption, the uniform distribution in Eq.(\ref{uni}) can be considered as a consequence of maximizing the Gibbs-Shannon entropy 
\be
H(\{ p_i \}) = -\sum_i p_i \ln p_i \label{GS}
\ee
under the constraint that the energy value $E_i$ of any microstate $i$ with nonzero probability should be $E$:
\be
\sum_i p_i (1 - \delta_{E_i E}) = 0 
\ee
along with the normalization condition
\be
\sum_i p_i  = 1.
\ee
These constraints are imposed using Lagrange multipliers, so that the function 
\bea
&&H(\{ p_i \}) + \lambda \sum_i p_i (1 - \delta_{E_i E})   + \nu (\sum_i p_i - 1)  \nonumber\\
&&= -\sum_i p_i \ln p_i  + \lambda \sum_i p_i (1 - \delta_{E_i E})  + \nu (\sum_i p_i - 1)
\eea
is to be varied with respect to $\{ p_i \},\lambda,\nu$ and then set to zero:
\bea
-\log p_i -1 + \lambda (1 - \delta_{E_i E}) + \nu &=& 0 \nonumber\\
\sum_i p_i (1 - \delta_{E_i E}) &=& 0 \nonumber\\
\sum_i p_i  - 1 &=& 0 \label{microconst}
\eea
The first line of Eq.(\ref{microconst}) yields 
\bea
p_i &=& e^{\nu-1} \quad (E_i = E) \nonumber\\
p_i &=& e^{\nu-1+\lambda} \quad (E_i \ne E),
\eea
and the second and the third line determine the values of the Lagrange multipliers, yielding\footnote{To avoid an infinite Lagrange multiplier, one may use $\tau$ defined as $\tau \equiv e^\lambda$ instead of $\lambda$.}
\bea
\nu &=& 1- \ln \Omega(E) \nonumber\\
\lambda &=& -\infty,
\eea
 resulting solution being the uniform distribution Eq.(\ref{uni}). The target function after the maximization is the Boltzmann entropy\footnote{In this work, the Boltzmann constant, which is just a unit conversion factor introduced for a historical reason, will be set to one, assuming that the temperature is measured in the same unit as the energy and the entropy is a dimensionless quantity.}
\be
S(E) = \ln \Omega(E). \label{bolzent}
\ee  
The maximization of the Gibbs-Shannon entropy (\ref{GS}) under macroscopic constraints  was termed as Maximum Entropy  principle by Jaynes\cite{jaynes57,jaynes03}  and proposed as the prescription  for obtaining the  equilibrium distribution, but the principle was applied mainly to open system, most probably because the uniform distribution that results for a close system  is considered as trivial. However, since the closed system can be considered as more fundamental and any result on open system can be derived as a special limit of a subsystem of a closed universe, it is natural to expect that the Maximum Entropy principle for an open system should be derivable from that for the closed one. 

Once the uniform distribution for a closed system is obtained, the Boltzmann distribution for a open system in thermal contact with a heat bath immediately follows, as can be found in any standard textbook on statistical physics\cite{PB,Path,Chan}. The open system and the heat bath form a closed universe, the probability distribution of microstates being the uniform one. The probability distribution of microstates for the open system only, which is a subsystem of the closed universe, is obtained as a marginal probability, where it is easy to see that the probability for a given microstate $i$ is proportional to the number of heat bath microstates consistent with the system energy $E_i$: 
\be p_i \propto \Omega_{\rm
bath}(E_{\rm tot}-E_i) = \exp\left[ S_{\rm bath} (E_{\rm tot}-E_i)\right] \simeq
\exp\left[S_{\rm bath} (E_{\rm tot})-\beta E_i\right] \label{Bol1} 
\ee where
$E_{\rm tot}$ is the total energy of the closed universe, which is a fixed constant, and the first order approximation in the last term comes from the fact that heat bath is  much larger than the open system so that the value of temperature $T = \left[\frac{d S_{\rm bath}}{d E_{\rm bath}}\right]^{-1}$ is a fixed constant that does not depend on the open system energy $E$. The standard assumption of weak coupling, that the energy is additive and the number states can be factorized for a given value $E_i$, is being used.   Since $S_{\rm bath} (E_{\rm
tot})$ is a constant, Eq.(\ref{Bol1}) is the Boltzmann distribution.

The same Boltzmann distribution can also be obtained by applying the Maximum Entropy principle directly to the open system, under the constraint that the expectation value of the fluctuating energy, $\sum p_i E_i$,  is fixed to a particular value $\epsilon$. Introducing the Lagrange multiplier, the function to extremize is now
\bea
&& H(\{ p_i \}) - \beta (\sum_i p_i  E_i - \epsilon )  + \nu (\sum_i p_i - 1)\nonumber\\
&&= -\sum_i p_i \ln p_i  - \beta (\sum_i p_i  E_i - \epsilon )  + \nu (\sum_i p_i - 1). \label{GS2}
\eea
The resulting solution takes the form of the Boltzmann distribution:
\be
p_i = \frac{e^{-\beta E_i}}{\sum_j e^{-\beta E_j}}.
\ee
where $\beta$ is determined by the condition
\be
\frac{\sum_k E_k e^{-\beta E_k}}{ \sum_i e^{-\beta E_i}} = \epsilon
\ee
However, the connection of the microcanonical derivation and the entropy maximization approach is not clear at all, since $\beta$ here is introduced as the Lagrange multiplier for constraining the mean energy, whereas $\beta$ appearing in the microcanonical derivation is the inverse of heat bath temperature, the derivative of the heat bath entropy with respect to energy. 

In the next section, it is shown that the Maximum Entropy principle for the open system is derived in the process of a two-step maximization of Gibbs-Shannon entropy for the entire closed universe, and the target function including the Lagrange term for constraining the energy expectation value of the open system is obtained by partial maximization of the Gibbs-Shannon entropy of the closed universe, over the microstate probability distribution of the closed universe.

\section{Maximum Entropy principle for open system, derived by partial  maximization of  the entropy of the closed universe}

In order to derive the target function of entropy maximization procedure for open system,  Eq.(\ref{GS2}), we first write down the Gibbs-Shannon entropy for the closed universe consisting of an open system and a heat bath, along with the Lagrange multiplier terms for the energy and normalization constraints:
\bea
&&H_{\rm tot}(\{p_{ia}\})+ \nu
(\sum_{ i,a} p_{ia} -1)  +  \lambda \sum_{i,a} p_{ia} (1 - \delta_{E_i+ E_a,  E}) \nonumber\\
&&= -\sum_{ i, a } p_{ia} \ln p_{ia} + \nu
(\sum_{ i,a} p_{ia} -1)  +  \lambda \sum_{i,a} p_{ia} (1 - \delta_{E_i+ E_a,  E})
\eea
where indices $i$ and $a$ labels the microstates of the open system and the heat bath, respectively. If the number of possible microstates in the open system and the heat bath are denoted as $A$ and $B$, we see there are $A B$ components of the variables $p_{ia}$, but the normalization condition $\sum_{ia} p_{ia}=1$ reduces the number of independent components to $AB -1$. We now change variables from $p_{ia}$ to $p_i \equiv \sum_a p_{ia}$ and $p(a | i) \equiv p_{ia}/p_i$. The number of components of    $p_i$ and $p(a | i)$ are $A$ and $AB$, but again the normalization conditions 
\bea
\sum p_i &=& 1 \nonumber\\
\sum_a p(a | i) &=& 1\quad (i= 1 \cdots A) \label{norm2}
\eea
reduce their number of independent components to $A-1$ and $AB-A$ respectively, making the total number of independent components to $AB-1$. Thus, to ensure the correct number of independent variables and maintain full equivalence, the normalization conditions (\ref{norm2}) should be implemented using the Lagrange multipliers. The target function for maximization, expressed in terms of the new variables, is:
\bea
&& -\sum_{ i, a } p(a | i) p_i \ln (p(a | i) p_i) + \sum_i \nu_i 
(\sum_{a} p(a | i) -1)  +  \mu (\sum_i p_i -1) + \lambda \sum_{i,a} p_i p(a | i )(1 - \delta_{E_i+ E_a,  E}) \nonumber\\
&&= -\sum_{ i} p_i \ln p_i -\sum_{ i, a } p(a | i) p_i \ln  p(a | i) + \sum_i \nu_i 
(\sum_{a} p(a | i) -1)  \nonumber\\
&&+  \mu (\sum_i p_i -1) + \lambda \sum_{i,a} p_i p(a | i ) (1 - \delta_{E_i+ E_a,  E}) \label{GSc}
\eea
where the conditions (\ref{norm2}) was used in producing the second line. 
Since the open system is the object of interest, we perform the maximization in two steps. First, the variation with respect to $p(a | i)$ , $\nu_i$, and $\lambda$ is performed, to eliminate them for given values of $p_i$. Then the target function is maximized with respect to the remaining variables  $p_i$ and $\mu$ at the second step. 

Taking variation of the target function in Eq.(\ref{GSc}) with respect to $p(a | i)$ , $\nu_i$, and $\lambda$ and setting them to zero, we get:
\bea
 -  p_i \ln  p(a | i) -p_i  + \nu_i + \lambda p_i  (1 - \delta_{E_i+ E_a,  E}) &=& 0 \label{var} \\
\sum_{a} p(a | i) &=& 1 \label{norm3}\\
 \sum_{i,a} p_i p(a | i ) (1 - \delta_{E_i+ E_a,  E})& = &0. \label{econst2}
\eea
From Eq.(\ref{var}) we get
\bea
p(a | i) &=& \exp(\frac{\nu_i}{p_i}-1) \quad (E_a = E_{\rm tot} - E_i) \nonumber \\
p(a | i) &=& \exp(\frac{\nu_i}{p_i}-1+\lambda) \quad (E_a \ne E_{\rm tot} - E_i).  \label{pai}
\eea
The constraints Eq.(\ref{econst2}) and Eq.(\ref{norm3}) fixes the values of $\lambda$ and $\nu_i$ so that we finally obtain
\be
p(a | i) = \frac{\delta_{E_a, E_{\rm tot} - E_i}}{\Omega(E_{\rm tot} - E_i)} \label{pai2}
\ee
Substituting Eq.(\ref{pai2}) into Eq.(\ref{GSc}), we now get
\bea
&&-\sum_{ i} p_i \ln p_i + \sum_{ i, a } p_i \frac{\delta_{E_a, E_{\rm tot} - E_i}}{\Omega(E_{\rm tot} - E_i)} \ln  \Omega(E_{\rm tot} - E_i)
+  \mu (\sum_i p_i -1)  \nonumber\\
&=& -\sum_{ i} p_i \ln p_i + \sum_{ i} p_i  \ln  \Omega(E_{\rm tot} - E_i)
+  \mu (\sum_i p_i -1) .
\eea
Now we use the fact that heat bath is much larger than the open system so that  $\ln  \Omega(E_{\rm tot} - E_i) \simeq \ln  \Omega(E_{\rm tot})- \frac{E_i}{T}$ to get
\be
\tilde H(\{ p_i \}) +  \mu (\sum_i p_i -1) \equiv -\sum_{ i} p_i \ln p_i  - \beta \sum_i p_i E_i +  \mu (\sum_i p_i -1). \label{opentarget}
\ee
where $\beta \equiv 1/T$ and an irrelevant constant term was dropped. Therefore, we see that the target function for  the Maximum Entropy  procedure for an open system, Eq.(\ref{GS2}) is now reproduced.  

This also shows that providing the open system with a thermal contact with a heat bath of fixed temperature, is equivalent to constraining the  expectation value of energy. To see this point more intuitively, note that the actual energy levels of the heat bath are not relevant, as long as the total energy of $E_{\rm tot}$ is maintained at the value satisfying
\be
[dS_{\rm bath}(E)/dE]_{E=E_{\rm tot}} = 1/T. \label{bathT}
\ee 
Then it is convenient to replace the heat bath with $N$ replica of the open system, where $N$ is to be taken to infinity. Maintaining the condition  Eq.(\ref{bathT}) then corresponds to fixing the average energy of $N+1$ open systems, $\bar E = E_{\rm tot}/N$. Since there are $N+1$ identical open systems, as $N \to \infty$, $\bar E$ becomes equal to the expectation value of energy for each open system $\langle E \rangle \equiv \sum_i p_i E_i$     due to the law of large numbers.
\section{Free energy minimization principle}
We  note that after multiplying $\tilde H(\{ p_i \};T)$ by $T$ and inverting the sign, we obtain
\be
\tilde F(\{ p_i \}; T) = \sum p_i E_i - T H (\{ p \}) =  \sum p_i E_i + T \sum_i p_i \ln p_i, 
\ee
which is to be minimized with respect to  $p_i$ under the normalization constraint $\sum_i p_i =1$. In fact, this function also appears in the literature\cite{Nel}, and since $\sum p_i E_i$ is the energy expectation value and $H(\{ p \})$ is an entropy, $\tilde F(\{ p_i \}; T)$ is considered as a Helmholtz free energy, and the corresponding minimization principle as the free energy minimization, but its microcanonical origin has never been explained so far. 

This is to be contrasted with the minimization of thermodynamic Helmholtz free energy,
\be
F(E; T) = E - T S (E) \label{free1}
\ee
which is a function of the open system energy $E$. The minimization  of Eq.(\ref{free1}) with respect to $E$ yields the most probable value of $E$ of the open system at equilibrium. On the other hand, the minimization of $\tilde F(\{ p_i \}; T)$ yields the equilibrium probability distribution of microstates.\footnote{Note  that  $F(E,T)$ and $\tilde F(\{ p_i \}; T)$ are distinct even after the minimizations. For the former, the minimum occurs at $E_{\rm min}(T)$ satisfying $dS/dE |_{E_{\rm min}(T)}= 1/T$ and the resulting function $A(T)=F(E_{\rm min}(T),T)$ is the Legendre transform of $S(E)$. On the other hand, the minimization of $\tilde F(\{ p_i \}; T)$ yields $\tilde A(T) = -T \ln \sum_i e^{-\beta E_i} = -T \ln \sum_E \exp(-\beta F(E; T))$. They are equal only in the limit of infinite system size.}

\section{Discussion}
There are two apparently distinct methods for deriving the Boltzmann distribution, one by considering the microcanonical ensemble of the closed universe consisting of the open system and the heat bath, and the other by maximizing the Gibbs-Shannon entropy of the open system under the constraint of fixed value of mean energy. In this work, it was shown that the Maximum Entropy procedure for the open system arises in the process of two-step maximization of the Gibbs-Shannon entropy of the closed universe, thus showing the equivalence of providing contact with the heat bath and constraining the energy expectation value. From this viewpoint, the microcanonical derivation and Maximum Entropy derivation of the Boltzmann distribution are the same thing. In the former, the Gibbs-Shannon entropy of the closed universe is maximized in one step, resulting in the uniform distribution of the microstate, which is then divided into open system and the heat bath to obtain the marginal probability distribution for the open system. In the latter, the maximization is performed in two steps, where partial maximization is performed over the microstates of the heat bath first, to obtain the target function expressed in terms of the probability distribution of the open system. Since it is a well-known fact that one-step or multi-step maximization of a function should yield the same result, it is no surprise that the same Boltzmann distribution is derived.

Although energy was the main focus in this work for the sake of concreteness, the argument can be generalized to any other macroscopic parameters such as volume or particle number. The entropy maximization procedure where the expectation value of  a macroscopic  parameter $\alpha$ is constrained, results in a Boltzmann form of distribution  $p_i \propto e^{-\gamma \alpha_i}$ with the corresponding Lagrange multiplier $\gamma$, and the microcanonical origin of this procedure is the entropy maximization for closed universe, consisting of the open system and the bath with fixed value of $\gamma=\partial S/\partial \alpha$.  

The case of particular interest is a dynamical system, where the probability distribution of dynamical paths is obtained by maximizing the Gibbs-Shannon entropy of path entropy, under the constraint on the expected value of quantities such as flux\cite{FK,jaynes85,Stock,philken,Steve11,HG,CecileJSM}. The argument of the present work provides an alternative justification, but it should be noted however, that the total ``universe"  consisting of the open system and the ``flux bath" cannot be interpreted as a closed universe, since a system with a nonzero flux is always a system driven by an external influence. In the case of dynamic system, one just considers  a large system with a fixed value of total flux.  Then, in the absence of any other additional information, the probability of each micropaths consistent with the constraint is equally probable, due to the Maximum Entropy principle. Then if we restrict our interest to the subsystem of this large system, the marginal probability distribution of the paths of the subsystem would be the one obtained by maximizing the path entropy under the constraint of fixed expectation value of the flux.  

\section{Acknowledgements}
{I thank Steve Press\'e, Kingshuk Ghosh, and Ken Dill for useful discussions.}

\end{document}